\documentclass[12pt]{article}

\input epsf
\def\hybrid{\topmargin 0pt      \oddsidemargin 0pt
        \headheight 0pt \headsep 0pt
        \voffset=-0.5cm
        \textwidth 6.25in       
        \textheight 9.5in       
        \marginparwidth 0.0in
        \parskip 5pt plus 1pt   \jot = 1.5ex}
\catcode`\@=11
\def\marginnote#1{}

\newcount\hour
\newcount\minute
\newtoks\amorpm
\hour=\time\divide\hour by60
\minute=\time{\multiply\hour by60 \global\advance\minute by-\hour}
\edef\standardtime{{\ifnum\hour<12 \global\amorpm={am}%
        \else\global\amorpm={pm}\advance\hour by-12 \fi
        \ifnum\hour=0 \hour=12 \fi
        \number\hour:\ifnum\minute<10 0\fi\number\minute\the\amorpm}}
\edef\militarytime{\number\hour:\ifnum\minute<10 0\fi\number\minute}

\def\draftlabel#1{{\@bsphack\if@filesw {\let\thepage\relax
   \xdef\@gtempa{\write\@auxout{\string
      \newlabel{#1}{{\@currentlabel}{\thepage}}}}}\@gtempa
   \if@nobreak \ifvmode\nobreak\fi\fi\fi\@esphack}
        \gdef\@eqnlabel{#1}}
\def\@eqnlabel{}
\def\@vacuum{}
\def\draftmarginnote#1{\marginpar{\raggedright\scriptsize\tt#1}}
\def\draftlabel#1{{\@bsphack\if@filesw {\let\thepage\relax
   \xdef\@gtempa{\write\@auxout{\string
      \newlabel{#1}{{\@currentlabel}{\thepage}}}}}\@gtempa
   \if@nobreak \ifvmode\nobreak\fi\fi\fi\@esphack}
        \gdef\@eqnlabel{#1}}
\def\@eqnlabel{}
\def\@vacuum{}
\def\draftmarginnote#1{\marginpar{\raggedright\scriptsize\tt#1}}

\def\draft{\oddsidemargin -.5truein
        \def\@oddfoot{\sl preliminary draft \hfil
        \rm\thepage\hfil\sl\today\quad\militarytime}
        \let\@evenfoot\@oddfoot \overfullrule 3pt
        \let\label=\draftlabel
        \let\marginnote=\draftmarginnote
   \def\@eqnnum{(\theequation)\rlap{\kern\marginparsep\tt\@eqnlabel}%
\global\let\@eqnlabel\@vacuum}  }

\def\numberbysection{\@addtoreset{equation}{section}
        \def\theequation{\thesection.\arabic{equation}}}

\def\underline#1{\relax\ifmmode\@@underline#1\else
        $\@@underline{\hbox{#1}}$\relax\fi}

\def\titlepage{\@restonecolfalse\if@twocolumn\@restonecoltrue\onecolumn
     \else \newpage \fi \thispagestyle{empty}\c@page\z@
        \def\thefootnote{\fnsymbol{footnote}} }

\def\endtitlepage{\if@restonecol\twocolumn \else  \fi
        \def\thefootnote{\arabic{footnote}}
        \setcounter{footnote}{0}}  
\relax

\hybrid

\newfont{\Bbb}{msbm10 scaled 1\@ptsize00}
\newcommand{\CC}{\mbox{\Bbb C}}

\newcommand{\ZZ}{\mbox{\Bbb Z}}
\newfont{\Bbbb}{msbm7 scaled 1\@ptsize00}
\newcommand{\z}{\raise-1pt\hbox{$\mbox{\Bbbb Z}$}}

\def\beq{\begin{equation}}
\def\eeq{\end{equation}}
\def\p{\partial}

\def\DD{{\sf D}}
\def\Dc{\CC \setminus {\sf D}}

\begin{document}
\begin{titlepage}

\title{Growth processes related to the dispersionless
Lax equations
\footnote{Based on the talk given at
the Workshop ``Physics and mathematics of growing interfaces",
Santa Fe, January 9-13, 2006}}

\author{A.~Zabrodin
\thanks{Institute of Biochemical Physics,
4 Kosygina st., 119991, Moscow, Russia
and ITEP, 25 B.Cheremushkinskaya, 117259,
Moscow, Russia}}

\date{September 2006}
\maketitle

\begin{abstract}
This paper is a short review of the connection
between certain types of growth processes and
the integrable systems theory, written from the
viewpoint of the latter.
Starting from the dispersionless Lax equations
for the 2D Toda hierarchy, we interpret them as evolution
equations for conformal maps in the plane.
This provides a unified approach to evolution of smooth
domains (such as Laplacian growth) and growth of
slits. We show that the L\"owner
differential equation for a parametric family of conformal maps
of slit domains arises as a consistency condition for reductions
of the dispersionless Toda hierarchy.
It is also demonstrated how the both types of growth
processes can be simulated by the large $N$ limit
of the Dyson gas picture for the model of normal random
matrices.

\end{abstract}

\vfill

\end{titlepage}

\section{Introduction}

Growth problems of Laplacian
type (such as Hele-Shaw viscous flows)
refer to dynamics of a moving front (an interface)
between two distinct phases
driven by a harmonic scalar field.
These essentially nonlinear
and non-local problems attract much attention
for quite a long time \cite{list,RMP}.

Remarkably,
Laplacian growth with vanishing surface tension possesses
an integrable structure uncovered in \cite{MWZWZ}.
Since evolution of planar domains is most naturally
described by time-dependent conformal maps, there is no
surprise that this structure is actually immanent for
general conformal maps and classical boundary value problems.
Specifically, in \cite{MWZ} it has been shown that evolution
of conformal maps is governed by an integrable
hierarchy of nonlinear partial differential equations which is
a zero dispersion version \cite{TakTak}
of the 2D Toda hierarchy \cite{UenoTakasaki}.
In fact the Lax equations for this hierarchy can be
derived from the classical theory of conformal maps depending
on parameters.

In the present paper we have tried to give a short
review of these and related developments in the context of the
integrable systems theory.
Taking the Lax representation of the
dispersionless 2D Toda (dToda) hierarchy as a starting point,
we follow how it induces the
contour dynamics and dynamics of conformal maps.
The Lax function, which is supposed to be univalent in some
neighborhood of infinity, is interpreted as a conformal map
from a fixed reference domain to
the complement of the growing domain.

Depending on the type of solutions to the dToda hierarchy, the
growing domains can be either smooth or singular like cuts or
slits. Hence the system of dispersionless Lax equations serves as
a master dynamical equation not only for Laplacian growth but also
for growth of slits. Conformal maps of parametric families of slit
domains are known to satisfy a differential equation proposed by
K.L\"owner in 1923 \cite{Lo}. Its connection with nonlinear
integrable equations was pointed out in \cite{GT}, see also
\cite{MMM,YuGibbons,TakTak1}. Following \cite{TTZ}, we demonstrate
how the radial L\"owner equation arises in the context of the
dToda hierarchy.

The unified treatment of the two
types of growth processes in the framework of the Toda integrable
system, which we emphasize in this paper, seems to be especially
promising in the light of the stochastic L\"owner evolution (SLE)
approach \cite{SLE}. This might give a hint how to incorporate, in
an intelligent way, a stochastic ingredient into growth problems
of Laplacian type.

An instructive representation of solutions to the dToda hierarchy
is provided by the large $N$ limit of certain matrix integrals
or their eigenvalue versions (the
Dyson gas representation), with
the associated growth processes being
simulated by evolution of support of eigenvalues. In
the last section we outline
the Dyson gas representation for Laplacian growth and growth of
slits.

Section 2 contains the necessary material on the dToda
hierarchy and its Lax representation. In section 3 we present
the general solution to the hierarchy in terms of canonical
transformations and distinguish the classes of
non-degenerate and degenerate solutions.
In section 4 we associate a contour dynamics with any solution
to the Lax equations. A particular subclass of degenerate
solutions is studied in section 5, where the L\"owner equation
is derived from the Lax equations.
Finally, section 6 contains the Dyson gas representation
for both non-degenerate and degenerate solutions.

\section{Lax representation for the dToda hierarchy}

This section contains a standard material which we present
in a form convenient for our purposes. For a more complete account
of dispersionless hierarchies, their algebraic structure, solutions
and applications see \cite{TakTak},\cite{KriW}-\cite{BMRWZ}.

\paragraph{Dispersionless Lax equations.}
We start with the Lax representation of the
dToda hierarchy with certain reality conditions imposed. The main
object is the Lax function $z(w)$ represented as a Laurent series of
the form
$$
z(w)= rw+ a_0 + \frac{a_1}{w}+\frac{a_2}{w^2} \, + \, \ldots
$$
The leading coefficient $r$ is assumed to be real while
all other coefficients $a_i$ are in general complex numbers.
All the coefficients depend on deformation
parameters (or ``times") $t_0$ (a real number)
and $t_1 , t_2 ,
t_3 , \ldots$ (complex numbers) in accordance with the
Lax equations
\beq\label{dtoda1}
\frac{\p z(w)}{\p t_k}=\{ A_k (w) , \, z(w)\}\,,
\quad
\frac{\p z(w)}{\p \bar t_k}=-\{ \bar A_k (w^{-1}) , \, z(w)\}
\eeq
where for any two functions of $w$, $t_0$
we set
\beq\label{bracket1}
\{ f, \, g\}
:=\frac{\p f}{\p \log w}\frac{\p g}{\p t_0}-
\frac{\p f}{\p t_0}\frac{\p g}{\p \log w}
\eeq
Here and below
the bar is complex conjugation and
$\bar f(w)$ means $\overline{f(\bar w)}$.
The reality condition thus
implies that the second half of the Lax equations (with
$\bar t_k$-derivatives) is obtained from the first one
by complex conjugation with $w$ on the unit circle.
The generators of the flows are constructed as follows:
$$
A_k(w)= \left ( z^k (w)\right )_{+}\,, \quad A_0(w)=\log w
$$
For the dToda hierarchy, the $(\ldots )_{+}$-operation
is
$$
\left ( z^k (w)\right )_{+}:=
\left ( z^k (w)\right )_{>0}+
\frac{1}{2}\left ( z^k (w)\right )_{0}
$$
Hereafter, $(\ldots )_{S}$ means taking the terms of the Laurent
series with degrees belonging to the subset $S\in \ZZ$ (in
particular, $(\ldots )_{0}$ is the free term). Note that at $k=0$
equations (\ref{dtoda1}) become tautological identities. The second
Lax function of the dToda hierarchy is $\bar z(w^{-1})$.
The reality conditions (i.e. the requirement that its coefficients are
complex conjugate to those of the $z(w)$) imply that
it obeys the same Lax equations.
It should be noted that (\ref{dtoda1}) is just a compact form of
writing evolution equations with respect to {\it real} deformation
parameters $t_{k}^{{\rm R}}={\cal R}e \, t_k$,
$t_{k}^{{\rm I}}={\cal I}m \, t_k$:
$$
\frac{\p z(w)}{\p t_{k}^{{\rm R}}}=\left \{
A_k(w)-\bar A_k (w^{-1}), \, z(w) \right \}\,, \quad
\frac{\p z(w)}{\p t_{k}^{{\rm I}}}=i\left \{
A_k(w)+\bar A_k (w^{-1}), \, z(w) \right \}
$$

We are especially interested in the class of solutions such that
$z(w)$, for all $t_k$ in an open set of the space of parameters, is
a univalent function in a neighborhood of infinity including the
exterior of the unit circle. This means that
in this neighborhood $z(w_1)=z(w_2)$ if and
only if $w_1 = w_2$. From now on, we assume that $z(w)$ belongs to
this class. In this case $z(w)$ is a conformal map from the exterior
of the unit circle to a domain in the complex plane containing
infinity while $\bar z(w^{-1})$ is a conformal map from the interior
of the unit circle to the complex conjugate domain.

Let $w(z)$ be the inverse function to the Lax function
$z(w)$. In terms of the inverse function, the evolution
equations (\ref{dtoda1}) acquire a simpler form:
\beq\label{dtoda2}
\frac{\p \log w(z)}{\p t_k}=\frac{\p A_k}{\p t_0}\,,
\quad
\frac{\p \log w(z)}{\p \bar t_k}=-\frac{\p \bar A_k}{\p t_0}
\eeq
Here $A_k= A_k (w(z))$, $\bar A_k =
\bar A_k (1/w(z))$ are regarded as functions
of $z$, and the derivatives are taken at constant $z$.

By purely algebraic manipulations, one can show \cite{TakTak} that
the compatibility conditions for the Lax equations (\ref{dtoda1})
read
$$
\p_{t_j}A_k (w)-\p_{t_k}A_j (w)+\{ A_k (w), A_j (w)\}=0
$$
$$
\p_{t_j}\bar A_k (w^{-1})+\p_{\bar t_k}A_j (w)+
\{ \bar A_k (w^{-1}), A_j (w)\}=0
$$
which is a dispersionless version of the ``zero curvature" representation.
Treating $A_k$'s as functions of $z$, one can rewrite them in the form
similar to (\ref{dtoda2}):
\beq\label{dtoda2a}
\frac{\p A_j}{\p t_k}=\frac{\p A_k}{\p t_j}\,,
\quad \frac{\p A_j}{\p \bar t_k}=-\frac{\p \bar A_k}{\p t_j}
\eeq
Note that at $j=0$ this system coincides with (\ref{dtoda2}).

\paragraph{Generating form of the Lax equations.}
Using a generating function of the polynomials $A_k$, the
infinite hierarchy (\ref{dtoda1}) can be represented as
a couple of ``generating equations". The generating function is
defined as
$$
\sum_{k\geq 1} \frac{z_{1}^{-k}}{k} A_k (w) =\sum_{k\geq 1}
\frac{\left ( z^k (w)\right )_{+}}{k z^k (w_1)}= -\left [ \log \left
(1-\frac{z(w)}{z(w_1)}\right ) \right ]_{+}
$$
where $w_1 = w(z_1)$. To separate the polynomial part and the free
term, we write
$$
\log \left (1-\frac{z(w)}{z(w_1)}\right )= \log \left
(1-\frac{w}{w(z_1)}\right ) + \log \frac{r w(z_1)}{z_1} + \log
\frac{z(w_1)-z(w)}{r(w_1 -w)}
$$
and notice that the expansion of the first (third) term contains
only positive (respectively, negative) powers of $w$ while the rest
is just the free term. Therefore,
\beq\label{dtoda3}
\sum_{k\geq 1}
\frac{z_{1}^{-k}}{k} A_k (w)= -\log (w(z_1 )-w) +\frac{1}{2}\log
\frac{z_1 w(z_1)}{r}
\eeq
Similarly,
\beq\label{dtoda3a}
\sum_{k\geq
1} \frac{\bar z_{1}^{-k}}{k} \bar A_k (w^{-1})= -\log (\bar w(\bar
z_1 )-w^{-1}) + \frac{1}{2}\log \frac{\bar z_1 \bar w(\bar z_1)}{r}
\eeq
and the generating Lax equations read
\beq\label{dtoda4}
D(z_1)
z(w)= -\left \{ \log (w(z_1 )-w) +\frac{1}{2}\log
\frac{r}{w(z_1)}\,, \, z(w) \right \}
\eeq
\beq\label{dtoda4a}
\bar D(\bar z_1) z(w)= \left \{ \log (\bar w(\bar z_1 )-w^{-1})
+\frac{1}{2}\log \frac{r}{\bar w(\bar z_1)}\,, \, z(w) \right \}
\eeq
where we have introduced the differential operators
$$
D(z)=\sum_{k\geq 1}\frac{z^{-k}}{k} \,  \p_{t_k }\,, \quad \bar
D(z)=\sum_{k\geq 1}\frac{z^{-k}}{k} \,  \p_{\bar t_k }
$$
Expanding both sides in powers of $z_1$, one recovers eqs.
(\ref{dtoda1}). In terms of the inverse function, the generating Lax
equations acquire a more transparent form (cf. (\ref{dtoda2})):
\beq\label{dtoda5}
D(z_1)\log w(z_2)=-\p_{t_0} \log \left (
w(z_1)-w(z_2)\right ) +\frac{1}{2}\p_{t_0}\log \frac{w(z_1)}{r}
\eeq
\beq\label{dtoda5a}
\bar D(\bar z_1)\log w(z_2)=\p_{t_0} \log \left
(\bar w (\bar z_1)-w^{-1}(z_2)\right )- \frac{1}{2}\p_{t_0}\log
\frac{\bar w(\bar z_1)}{r}
\eeq
Tending $z_2 \to \infty$, we obtain
the useful equations
\beq\label{dtoda6}
D(z)\log r
=-\frac{1}{2}\p_{t_0} \log \left ( rw(z)\right )\,,
\eeq
\beq\label{dtoda6a}
\bar D(\bar z)\log r =-\frac{1}{2}\p_{t_0} \log
\left ( r\bar w(\bar z)\right )
\eeq
Plugging them back into eqs.
(\ref{dtoda5}), (\ref{dtoda5a}), one can represent the latter
relations in a slightly more compact form:
\beq\label{dtoda7}
D(z_1)\log (rw(z_2))=-\p_{t_0} \log \left ( rw(z_1)-rw(z_2)\right )
\eeq
\beq\label{dtoda7a}
\bar D(\bar z_1)\log (rw(z_2))=\p_{t_0}
\log \left (1-\frac{1}{\bar w(\bar z_1)w(z_2)}\right )
\eeq

The construction of $A_k$'s implies that the expansion of
$A_k (w(z))$ in a Laurent series in $z$ looks like
$A_k = z^k + O(1)$. Furthermore, the compatibility conditions
(\ref{dtoda2a}) and relation (\ref{dtoda3}) allow one to represent
the coefficients in the form
\beq\label{Ak0}
A_0 (w(z))=\log w(z) =  -\frac{1}{2}\, \p_{t_0}v_0 -\sum_{k\geq 1}
\frac{\p_{t_0}v_k}{k}\, z^{-k}
\eeq
\beq\label{Ak1}
A_j(w(z))=z^j -\frac{1}{2}\, \p_{t_j}v_0 -\sum_{k\geq 1}
\frac{\p_{t_j}v_k}{k}\, z^{-k} \,, \quad j\geq 1
\eeq
\beq\label{Ak1a}
\bar A_j(w^{-1}(z))=\frac{1}{2}\, \p_{\bar t_j}v_0 +\sum_{k\geq 1}
\frac{\p_{\bar t_j}v_k}{k}\, z^{-k} \,, \quad j\geq 1
\eeq
where $v_k$ are functions of the times such that
$\p_{t_j}v_k=\p_{t_k}v_j$, $\p_{t_j}\bar v_k=\p_{\bar t_k}v_j$.
The latter conditions allow one to introduce a real-valued function
${\cal F}$ via $v_k = \p_{t_k}{\cal F}$. Equations (\ref{dtoda7}),
(\ref{dtoda7a}) then become the
dispersionless Hirota equations for the function ${\cal F}$.

\section{General solution to the Lax equations}

A general solution to the differential equations (\ref{dtoda1}) is
available in an implicit form \cite{TakTak}.
To present it, we need an extended
version of the Lax formalism.

The idea is as follows. By the definition of the Poisson bracket,
$\log w$ and $t_0$ form a canonical pair: $\{\log w, \, t_0\}=1$.
The evolution according to the Lax equations can be regarded as a
$t_k$-dependent canonical transformation from the pair $(\log w,
t_0)$ to another canonical pair whose first member is $\log z(w)$.
It is quite natural to introduce the second member which we denote
by $M$. Depending on the situation, we shall treat it either as a
function of $z$ and $t_0$ or as a function of $w$ and $t_0$ through
the composition $M=M(z(w,t_0), t_0)$ (it also depends on the
deformation parameters $t_k$). To find what is $M$, we note that the
condition $\{\log z, \, M\}=1$ can be identically rewritten as
$\p_{t_0}M(z)=z\p_z \log w(z, t_0)$. This determines $M$ up to a
term depending only on $z$. The latter is fixed if one requires $M$
to obey the same Lax equations (\ref{dtoda1}). To wit, equation
$\p_{t_k}M = \{ A_k,  M\}$ (where the derivatives are taken at
constant $w$) is equivalent to
$$
\p_{t_k}M(z)=w\p_w A_k \, \p_{t_0}M(z) = z \p_z A_k
$$
Taking into account (\ref{Ak0}), (\ref{Ak1}), we can write
\beq\label{M1}
M=\sum_{k\geq 1}kt_kz^k (w) + t_0  + \sum_{k\geq 1}v_k z^{-k}(w)
\eeq
It is the quasiclassical (dispersionless) limit of the
Orlov-Shulman operator \cite{OS}. Its geometric meaning depends on
the choice of a particular solution. In a similar way, one can
construct the conjugate Orlov-Shulman function, $\bar M(\bar z)$,
such that the transformation $(\log w,  t_0 ) \rightarrow (\log \bar
z^{-1}(w^{-1}), \bar M (\bar z (w^{-1}))$ is canonical and $\bar M$
obeys the same Lax equations. The Lax equations imply that the
composition of the canonical transformations
$$
(\log z, M) \rightarrow (\log w, t_0 ) \rightarrow (\log \bar
z^{-1}, \bar M)
$$
{\it does not depend on $t_k$}, i.e., it is an integral of motion.
Moreover, any $t_k$-independent canonical transformation $(\log z,
M) \rightarrow (\log \bar z^{-1}, \bar M)$
between the Laurent series of the form prescribed above
generates a solution to
the dToda hierarchy. A detailed proof can be found in \cite{TakTak}.

More precisely, let $(\log f (w, t_0), g(w, t_0))$
be a canonical pair: $\{\log f, g\}=1$.
Suppose that the functions $z, \bar z, M, \bar M$ of the
form as above are connected by the functional relations
\beq\label{M2}
1/\bar z (w^{-1})=f\left (z(w), M(z(w) )\right )\,, \quad \bar
M(\bar z(w^{-1}))=g\left (z(w), M(z(w))\right )
\eeq
Then the function $z(w)$
obeys the hierarchy of the Lax equations and its coefficients (as
functions of $t_k$'s) thus obey the dToda hierarchy. Conversely, any
solution of the dToda hierarchy admits a representation
of this form with some $(f, g)$-pair. Note that the reality
conditions imply the following constraints on the functions
$f$, $g$:
\beq\label{reality}
\bar f^{-1}\left (f^{-1}(w, t_0), \, g(w,t_0)\right ) =w \,,
\quad
\bar g\left (f^{-1}(w, t_0), \, g(w,t_0)\right ) =t_0
\eeq
which is a sort of the ``unitarity condition" for the canonical
transformation $(\log w, t_0)\rightarrow (\log f(w, t_0), g(w, t_0))$.

This construction can be made more explicit by introducing the
generating function of the canonical transformation $(\log w ,
t_0)\rightarrow (\log f, g)$. An important class of solutions
corresponds to the canonical transformations $(\log z, M)\rightarrow
(\log \bar z^{-1}, \bar M)$
defined by means of a generating function $U(z, \bar z)$
\cite{Ztmf}:
\beq\label{M3}
M=z\p_z U(z, \bar z)\,,
\quad \bar M=\bar z\p_{\bar z} U(z, \bar z)
\eeq
Here $U(z, \bar z)$
can be an arbitrary differentiable real-valued function of
$z$, $\bar z$. This form of the canonical transformation implies
that the functions $z(w)$ and $\bar z(w^{-1})$ are algebraically
independent. (By
algebraic dependence we mean here existence of a
$t_k$-independent function of two variables $R(z, \bar z)$ such that
$R(z(w), \bar z(w^{-1}))=0$ for all $t_k$.) These are
solutions of generic type.
We call them {\it non-degenerate}.
For non-degenerate solutions the ``string equation"
\beq\label{M4}
\{z(w), \, \bar z(w^{-1}) \}=\frac{1}{U_{z \bar z}(z(w), \bar z(w^{-1}))}
\eeq
where $U_{z \bar z} (z, \bar z)\equiv \p_z \p_{\bar z}U(z, \bar z)$
holds true. It is obtained by plugging $M$ from (\ref{M3}) into the
canonical relation $\{z, M\}=z$.

The origin of the string equation can be
understood in a simpler way as follows. From the
Lax equations (\ref{dtoda1})
and the Jacobi identity for the Poisson bracket it follows that
$\{z(w), \bar z(w^{-1})\}$ obeys the same Lax equations:
$$
\p_{t_k} \{z(w), \bar z(w^{-1})\}=\left \{ A_k(w),
\, \{z(w), \bar z(w^{-1})\}\right \}
$$
Therefore, any relation of
the form
\beq\label{M4a}
\{z(w), \bar z(w^{-1})\}=\omega (z(w), \bar z(w^{-1}))
\eeq
where $\omega (z, \bar z)$ is an arbitrary $t_k$-independent
function of two variables such that $\bar \omega (z, \bar z)=\omega
(\bar z, z)$ is consistent with the hierarchy. The approach based on
the canonical transformations clarifies the meaning of this function
and makes it clear that in fact any solution obeys a string equation
of the form (\ref{M4a}).

Canonical transformations such that the
functions $z(w)$ and $\bar z(w^{-1})$
appear to be algebraically
dependent can not be represented
in the form (\ref{M3}).
They correspond to solutions which we call
{\it degenerate}. For degenerate solutions,
the Poisson bracket $\{z, \bar z\}$ vanishes. Conversely,
the relation $\{z, \bar z\}=0$ implies the algebraic dependence.
Indeed, in terms of the function $\bar z(w^{-1}(z))$
the Lax equation for $\bar z(w^{-1})$ reads:
$$
\p_{t_k}\bar z(w^{-1}(z))=\p_z A_k \,\{z, \bar z\}
$$
(the derivatives are taken at constant $z$),
so at $\{z, \bar z\}=0$ we have $\p_{t_k}\bar z(w^{-1}(z))=0$
for all $t_k$.
This just means that
the relation between $z(w)$ and $\bar z(w^{-1})$ is
$t_k$-independent.
A particular subclass of
degenerate solutions, together with their geometric interpretation,
is discussed below in section 5.

\section{Contour dynamics}

\begin{figure}[tb]
\epsfysize=4.5cm
\centerline{\epsfbox{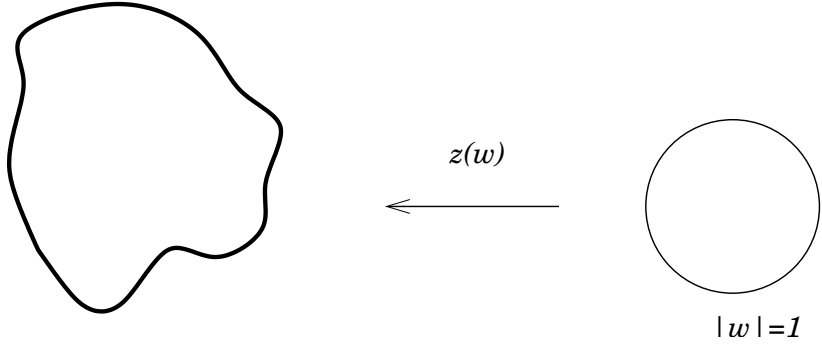}}
\caption{\sl The Lax contour. }
\label{fi:lax.eps}
\end{figure}

The Lax equations (\ref{dtoda1}) can be understood as equations of a
contour dynamics. The contour is the image of the unit circle, i.e.,
$z(e^{i\theta})$, $0\leq \theta \leq 2\pi$. Let us call it the Lax
contour (Fig.~\ref{fi:lax.eps}) and denote it by $\gamma$.
It depends on the deformation parameters according to the
Lax equations.

We need a general kinematic relation. Let $(x(\sigma , t), y(\sigma,
t))$ be any parameterizations of a moving contour in the plane, then
the normal velocity of the contour points is
$$
V_n = \frac{d\sigma }{dl} \left ( \p_{\sigma}x \p_t y -
\p_{\sigma}y \p_t x\right )
$$
where $dl =\sqrt{(dx)^2 +(dy)^2}$ is the line element along the contour.

Applying this formula to the Lax contour $z(e^{i\theta})$ with the
specific parametrization $\sigma =\theta$ and $t=t_0$ with all other
$t_k$'s fixed, we get the normal velocity of the Lax contour $\gamma$
at the points
$z(w)$, $|w|=1$:
\beq\label{vn}
V_n = -\, \frac{\{z(w), \, \bar z (w^{-1})\}}{2|z'(w)|}
\eeq
Here $z'(w)=\p_w z(w)$ and the Poisson bracket in the numerator
is given by (\ref{bracket1}).
More generally, the normal velocities corresponding to the changes
of the (real) times $t_{k}^{{\rm R}}={\cal R}e \, t_k$,
$t_{k}^{{\rm I}}={\cal I}m \, t_k$ are given by
\beq\label{trk}
V_{n}^{(t_{k}^{{\rm R}})}=
-\, \frac{\{z(w), \, \bar z (w^{-1})\}}{2|z'(w)|}\,
\left (\phi_k (w)+ \bar \phi_k (w^{-1})\right )
\eeq
\beq\label{tik}
V_{n}^{(t_{k}^{{\rm I}})}= -\, i \,\frac{\{z(w), \, \bar
z (w^{-1})\}}{2|z'(w)|}\,
\left (\phi_k (w)- \bar \phi_k
(w^{-1})\right )
\eeq
where
\beq\label{phi}
\phi_k (w) :=w\p_w A_k (w)
\eeq
The function $z(w)$ provides a time-dependent conformal map
from the exterior of the unit circle onto the exterior of the
Lax contour.

\paragraph{Non-degenerate solutions.}
The non-degenerate solutions corresponding to the canonical
transformation with the generating function $U(z, \bar z)$
(\ref{M3}) have a clear interpretation in terms of contour dynamics.
Eq. (\ref{vn}) together with the string equation (\ref{M4}) states
that the normal velocity of the Lax contour at the point $z\in \gamma$
is equal to
\beq\label{vn2}
V_n (z)=  - \,
\frac{|w'(z)|}{2\p_z \p_{\bar z}U(z, \bar z)} \,, \quad z\in \gamma
\eeq
Eqs. (\ref{M1}), (\ref{M3}) allow us to express the deformation parameters
in terms of the moving contour:
\beq\label{vn3}
t_k=\frac{1}{2\pi ik}\oint_{|w|=1}z^{-k-1}(w)M(z(w)) dz(w)=
\frac{1}{2\pi ik}\oint_{\gamma} z^{-k}\p_z U \, dz \,,
\quad k\geq 1
\eeq
\beq\label{vn3a}
t_0=\frac{1}{2\pi i}\oint_{|w|=1}M(z(w)) d\log z(w)=
\frac{1}{2\pi ik}\oint_{\gamma} \p_z U \, dz
\eeq
We stress that $t_1, t_2, \ldots$ are kept constant, so they
are integrals of motion for the contour dynamics (\ref{vn2}).

A particularly
important case is $U(z,\bar z)=z\bar z$ which corresponds to the
canonical transformation $\bar z = z^{-1}M$, $\bar M = M$ (i.e.,
$M=\bar M =z\bar z$). In this case the normal velocity
is given by
\beq\label{vn4}
V_n (z)=  - \,
\frac{1}{2} |w'(z)|  \,, \quad z\in \gamma
\eeq
Note that $|w'(z)|$ is equal to
the normal derivative $\p_n \log |w(z)|$ of the solution to the
Laplace equation with a source at infinity and the Dirichlet boundary
condition on the contour.
We thus see that (\ref{vn4}) is identical to
the Darcy law for the dynamics of interface between
viscous and non-viscous fluids confined in the radial Hele-Shaw cell,
assuming vanishing surface tension on the interface. Formulas (\ref{vn3})
state that $t_k=\frac{1}{2\pi ik}\oint_{\gamma} z^{-k}\bar z dz$ are harmonic
moments of the exterior of the contour $\gamma$.
Their conservation in the course of the Laplacian growth dynamics was
first established by S.Richardson \cite{Richardson}.
Eq. (\ref{vn3}) states that the time variable $t_0$
should be identified with area (divided
by $\pi$) of the
interior domain encircled by $\gamma$.

The Laplacian growth with the source at a finite point
$z_0$ corresponds to the same function $U(z, \bar z)=z\bar z$
and the vector field
$\p_{t_0} + D(z_0)+\bar D(\bar z_0)$ in the space of
deformation parameters. Indeed, using eqs. (\ref{trk}), (\ref{tik})
and relations (\ref{dtoda3}), (\ref{dtoda3a}) we obtain the
normal velocity
\beq\label{vn5}
V_n (z)=   \frac{\p_n G(z,z_0)}{2\p_z \p_{\bar z}
U(z, \bar z)} \,, \quad z\in \gamma
\eeq
where
$$
G(z,z_0)=\log \left |
\frac{w(z)-w(z_0)}{1-w(z)\overline{w(z_0)}}\right |
$$
is the Green function of the Dirichlet boundary value problem
in the exterior of the Lax contour.

\begin{figure}[tb]
\epsfysize=4.5cm
\centerline{\epsfbox{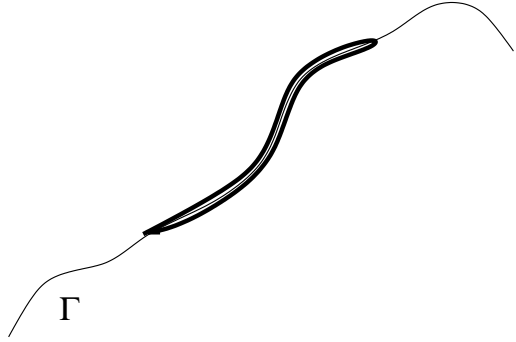}}
\caption{\sl A degenerate Lax contour. The arc can move along
a fixed curve $\Gamma$.}
\label{fi:arc}
\end{figure}

\paragraph{Degenerate solutions.}
Degenerate solutions describe evolution
of singular contours like growth of slits or cuts in the plane.
Since the Poisson bracket $\{z, \bar z\}$ vanishes, it might seem
from (\ref{vn}) that the velocity of the contour vanishes as well
and so there is no dynamics at all. In fact this is not exactly the case:
$V_n$ does vanish unless $z'(w)=0$.
We see that the growth is possible only
at the points that are images of the critical points of the conformal map
$z(w)$ lying on the boundary (on the unit circle). This means that
only endpoints of arcs can move while other boundary points remain fixed.
We see that
the Lax contour or at least a finite part of it degenerates
into an arc of a fixed curve $\Gamma$ swept twice (back and forth),
or into a collection of such arcs, and the evolution consists in
moving the endpoints of the arc along the same curve $\Gamma$
(see an example in Fig.~\ref{fi:arc}).
This agrees with the fact that for degenerate solutions the
functions $z(w)$ and $\bar z(w^{-1})$ are algebraically dependent:
the $t_k$-independent relation $R(z(w), \bar z(w^{-1}))=0$ is just
the equation of the curve $\Gamma$.
The deformation parameters $t_k$ do not admit so transparent
interpretation as in the non-degenerate case.

\section{Reductions of rank 1 and
radial L\"owner equation}

Any particular solution of the dToda hierarchy can be
regarded as a transition from the variables $t_0, t_1, t_2, \ldots$
to the variables $r, a_0, a_1, \ldots$ which are coefficients
of the Lax function $z(w)$: $z(w)=rw +a_0 + a_1 z^{-1}+\ldots$.
The non-degenerate solutions result in true changes
of variables, i.e., the Jacobian of this transition does not vanish.
For degenerate solutions, the Jacobi matrix is degenerate.
In this section we present a detailed analysis of the simplest
nontrivial case when this matrix is of rank 1.
We say that the corresponding
solutions are {\it reductions of rank 1} (of the dToda hierarchy).
As is easy to see,
the reduction of rank 1 implies that the Lax function and thus
$w(z)$ depends on all the times $t_j$ through {\it only one independent
function} $q=q(\{t_j \})$:
\beq\label{rloe1}
z(w; \{t_j \})=z(w,q)\,,
\quad w(z; \{t_j \})=w(z,q)
\eeq
Our goal is to characterize the
form of the function $w(z)=w(z,q)$
or $z(w)=z(w,q)$
consistent with the dToda hierarchy.
Without loss of generality, we set $q:=\log r$.
We shall see that the consistency condition is the
{\it radial L\"owner equation} \cite{Lo}
\beq\label{rLoewner}
\frac{\p w(z)}{\p \, q}=w(z)\,
\frac{\eta (q)+w(z)}{\eta (q)-w(z)}
\eeq
where $\eta (q)$ is arbitrary continuous function of $q$
such that $|\eta (q)|=1$ (the ``driving function").
This equation is well known
in the theory of univalent functions (see, e.g., \cite{univalent})
as a differential equation for conformal maps of slit domains
parameterized by a parameter $q$.

In the calculation below, we closely follow \cite{TTZ}.
Let us plug the ansatz
(\ref{rloe1}) into the equation (\ref{dtoda7}).
Using the chain rule of differentiation, we have:
$$
D(z_1)q \cdot \p_q \log (e^q w(z_2))= -\,
\frac{\p_q (e^q w(z_1))-\p_q (e^q w(z_2))}{e^q (w(z_1)-w(z_2))}
\, \p_{t_0}q
$$
Now, using (\ref{dtoda6}) and assuming that
$\p_{t_0}q \neq 0$, we get the relation
\beq\label{rloe2}
\frac{1}{2} \p_q \log (e^q w(z_1)) \, \p_q \log (e^q w(z_2)) =
\frac{w(z_1)\p_q \log (e^q w(z_1))-
w(z_2)\p_q \log (e^q w(z_2))}{w(z_1)-w(z_2)}
\eeq
which means that the combination
\beq\label{etaq}
\eta (q)=- w(z)\frac{1+\p_q \log w(z)}{1-\p_q \log w(z)}
\eeq
does not depend on $z$.
This implies the radial L\"owner
equation (\ref{rLoewner})
or, for the Lax function $z(w,q)$,
\beq\label{rLoewner1}
\frac{\p z(w)}{\p \, q}=-\, w\,
\frac{\eta (q)+w}{\eta (q)-w}\,\,
\frac{\p z(w)}{\p \, w}
\eeq
Note that the functions $1/\bar w(z)$ and $\bar z(w^{-1})$ obey
the same L\"owner equations (\ref{rLoewner}) and (\ref{rLoewner1})
respectively.

Equation (\ref{dtoda7a}) implies that $|\eta (q)|=1$. Indeed, with
the reduction imposed it becomes
$$
\frac{1}{2}\, \p_{q}\log (e^{q}\bar w(\bar z_1))\, \p_{q}\log
(e^{q} w(z_2))= \frac{\p_{q}\log (\bar w(\bar z_1)
w(z_2))}{1-\bar w(\bar z_1)w(z_2)}
$$
or, after rearranging,
$$
\bar w(\bar z_1)\, \frac{1+\p_{q}\log
\bar w(\bar z_1)}{1-\p_{q}\log  \bar w(\bar z_1)}\,
w(z_2)\, \frac{1+\p_{q}\log
w(z_2)}{1-\p_{q}\log  w(z_2)} =1
$$
that just means that $\overline{\eta (q)}\eta (q)=1$.
In fact this constraint follows already from eq. (\ref{etaq}):
let $z$ belong to the image of the unit circle
under the map $z(w)$ (i.e., to the Lax contour), then complex conjugation
of (\ref{etaq}) yields $\overline{\eta (q)}=\eta^{-1}(q)$.

Using the L\"owner equations for $z(w)$ and $\bar z(w^{-1})$, it is
straightforward to verify that $\{z(w), \bar z(w^{-1})\}=0$. As it
was argued in section 3, this means a $t_k$-independent relation
$R(z, \bar z)=0$ between the Lax functions $z$ and $\bar z$ which
defines a curve $\Gamma$ in the plane. Let us represent it in the form
$\bar z = S_{\Gamma}(z)$. The function $S_{\Gamma}$ is called the
Schwarz function of the curve $\Gamma$ \cite{Davis}.
For the degenerate solution of rank 1 under consideration
it is an integral of motion.
It is easy to see that the canonical
transformation (\ref{M2}) corresponding to this solution
can be written in terms of the Schwarz function as follows:
\beq\label{rloe5}
\bar z=  1/S_{\Gamma}(z) \,, \quad \bar M =
- \, \frac{S_{\Gamma}(z)}{zS'_{\Gamma}(z)}\, M
\eeq
The reality constraint (\ref{reality}) follows from the identity
$\bar S_{\Gamma}(S_{\Gamma}(z))=z$ obeyed by the Schwarz function.

As $w$ sweeps the unit circle, $z(w)$ sweeps an arc
of the curve $\Gamma$ (back and forth).
The arc depends on all the times
through $q$. The function $z(w)$ conformally maps the exterior
of the unit circle onto the complement of the arc. We see that
this map does obey the L\"owner equation as it must.

The dependence of $q$ on the times $t_k$, $\bar t_k$
is determined by a system of equations of hydrodynamic type.
They follow from eq. (\ref{dtoda6})
which can be written as
$D(z)q
=-\,\frac{1}{2}(1+\p_{q}\log w(z))\p_{t_0}q$.
Using the L\"owner equation, we obtain:
\beq\label{rloe3}
D(z)q
=\frac{\eta (q )}{w(z)-\eta (q)} \, \p_{t_0}q
\eeq From (\ref{dtoda3}) we conclude that
$$
\frac{\eta }{w(z)-\eta}= \sum_{k\geq 1}\frac{z^{-k}}{k}\,
\phi_{k}(\eta)
$$
with $\phi_k (w)$ as in (\ref{phi})
and thus the system of equations of hydrodynamic type reads
\beq\label{rloe4}
\p_{t_k}q=\phi_k (\eta (q)) \p_{t_0}q\,, \quad k=1,2,\ldots
\eeq
Equations containing $\bar t_k$-derivatives are
obtained by complex conjugation.

At last, it should be mentioned that the chordal version
of the L\"owner equation (see, e.g., \cite{SLE}), emerges,
in a similar way, in the context of the dispersionless
KP hierarchy \cite{GT,MMM,YuGibbons}.

\section{The large $N$ Dyson gas representation of solutions
to the dToda hierarchy}

In this section, we reconstruct the solutions of the dToda
hierarchy (both non-degenerate and degenerate) using the
Dyson gas representation, i.e., eigenvalue
versions of matrix integrals for certain models of random matrices.

Consider the following $N$-fold integral over the complex plane:
\beq\label{tauN}
\tau_N = \frac{1}{N!} \int_{\CC} \prod_{m<n}|z_m - z_n|^2
\prod_{j=1}^{N} e^{\frac{1}{\hbar} \sum_{k\geq 1} (t_k
z_{j}^{k}+\bar t_k \bar z_{j}^{k})} \, d\mu (z_j , \bar z_j)
\eeq
where $d\mu$ is some integration measure
and $\hbar$ is a parameter. For
$d\mu = e^{-\frac{1}{\hbar}U(z, \bar z)} d^2z$
the integral is equal to the partition
function of the model of normal random matrices with the potential
$2{\cal R}e \, \sum_k t_k z^k -U(z, \bar z)$
written as an integral over eigenvalues.
Equivalently, it is equal to the partition function of
the system of 2D Coulomb charges interacting via the logarithmic
potential in an external field (the Dyson gas).
It appears that both Laplacian growth and growth of slit domains
can be simulated by the large $N$ limit of this integral.

The basic fact linking
the integral (\ref{tauN}) to integrable systems is that
for any measure $d\mu$ (including singular measures
supported on sets of dimension less than $2$),
$\tau_N$, as a function of $\{t_k \}$,  $\{\bar t_k \}$,
is a $\tau$-function of the 2D Toda hierarchy with a
nonzero dispersion parameter proportional to $\hbar$, i.e.,
it obeys the full set of Hirota bilinear identities
for this hierarchy
\cite{Sato,UenoTakasaki}.
In a slightly different form, this statement
first appeared in \cite{KMMM}, see also
\cite{Zabor}.
The dispersionless version is reproduced
in the large $N$ limit such that $N\to \infty$, $\hbar \to 0$,
and $t_0 = \hbar N$ remains finite.
Then $\tau_N$ generates the dispersionless
``$\tau$-function" (or rather ``free energy" ) ${\cal F}$ via
\beq\label{Ftau}
{\cal F}(t_0 , \{t_k \}, \{\bar t_k \})=
\lim_{N\to \infty} \left (
\hbar^2 \log \tau_N \right)
\eeq
It obeys the dispersionless Hirota relations (see below).

Second order $t_k$-derivatives of ${\cal F}$ enjoy a nice
geometric interpretation through conformal maps.
This goes as follows. As $N \to \infty$,
the integral (\ref{tauN}) is determined by the
most favorable configuration of
$z_i$'s, i.e., the one at which the integrand has
a maximum. Using the electrostatic analogy, one can see
that this holds when the points $z_i$ (2D Coulomb charges)
densely fill a bounded domain $\DD$ in the complex plane.
In terms of the mean density of the charges
$\left < \rho (z)\right >=
\hbar \left < \sum_k \delta^{(2)}(z-z_k)\right >$
this domain is characterized by the condition
$$
\lim_{N\to \infty} \left < \rho (z)\right > >0 \quad
\mbox{if $z\in \DD$} \quad \mbox{and} \quad
\lim_{N\to \infty} \left < \rho (z)\right > =0 \quad
\mbox{otherwise}
$$
For simplicity, we assume that $\DD$ is connected. In the
matrix model interpretation, this domain is called the
support of eigenvalues.

Let $w(z)$ be the conformal mapping function from
the exterior
of the domain $\DD$ onto the exterior of the unit circle
normalized as $w(z)=z/r +O(1)$ at large $|z|$ with a real
$r$ called the exterior conformal radius of the domain $\DD$.
In \cite{KKMWZ,Ztmf} it was shown that the function
$w(z)$ can be expressed through ${\cal F}$ in the following
different but equivalent ways:
\beq\label{confmap1}
rw(z)= z\, e^{-D(z)\p_{t_0}{\cal F}}
\eeq
\beq\label{confmap2}
rw(z)=z- a -D(z)\p_{t_1}{\cal F}
\eeq
\beq\label{confmap3}
rw^{-1}(z)= D(z) \p_{\bar t_1}{\cal F}
\eeq
where
\beq\label{r1}
2\log r =\frac{\p^2  {\cal F}}{\p t_{0}^{2}}\,,
\quad
a =\frac{\p^2  {\cal F}}{\p t_{0} \p t_1}
\eeq
and the operator $D(z)$ is defined in section 2.
The consistency of these relations follows from equations of the
dToda hierarchy.
The 2D dToda hierarchy can be written in a generating
form as
\beq\label{Toda1}
D(z_1)D(z_2){\cal F}=\log
\frac{rw(z_1)-rw(z_2)}{z_1 -z_2}
\eeq
\beq\label{Toda3}
-  D(z_1)\bar D(z_2){\cal F}=\log
\left ( 1-\frac{1}{ w(z_1)\bar w(z_2)}\right )
\eeq
together with complex conjugate equations ( cf. eqs.
(\ref{dtoda7}), (\ref{dtoda7a})
which are $t_0$-derivatives of (\ref{Toda1}), (\ref{Toda3})).

We emphasize that all the relations given above
hold true for {\it any} measure $d\mu$ in (\ref{tauN})
provided the most favorable configuration of $z_i$'s at
$N \to \infty$ is well defined. If the measure is smooth, say
$d\mu = e^{-\frac{1}{\hbar}\, U(z, \bar z)} d^2z$, then
this construction gives non-degenerate solutions to the dToda
hierarchy discussed in section 3.
It is easy to see that $U(z, \bar z)$ is just the
generating function of the canonical transformation (\ref{M3}),
hence the notation.
Singular measures $d\mu$ lead to degenerate solutions.
In particular, one may consider the measure supported
on a curve $\Gamma$, then the integral (\ref{tauN}) becomes
one-dimensional (along $\Gamma$) in each variable:
\beq\label{tauN1}
\tau_N = \frac{1}{N!} \int_{\Gamma} \prod_{m<n}|z_m - z_n|^2
\prod_{j=1}^{N} e^{\frac{1}{\hbar}
\sum_{k\geq 1} (t_k z_{j}^{k}+\bar t_k \bar z_{j}^{k})}
\, |dz_j|
\eeq
In the large $N$ limit, the support of eigenvalues, $\DD$,
is then an arc of the curve $\Gamma$ (or several disconnected
arcs). The function $w(z)$ maps the slit domain $\Dc$ onto the exterior
of the unit circle.
The choice of the measure
supported on a curve
means a reduction of the dToda hierarchy.
A familiar example is the dToda chain,
where one may take $\Gamma$ to be either real or imaginary axis.
Consider a general (continuous)
curve $\Gamma$ infinite in both directions.
It is clear that
$w(z)$ and the Lax functions (the functions inverse to
$w(z)$ and $\bar w(z)$) depend on the times through two
parameters only. One can set them to be, for example,
the positions of the two ends of the arc $\DD$ on the curve
$\Gamma$.
This is a reduction of rank 2.
The simplest way to obtain a reduction of rank 1 is
to take the measure $d\mu$
supported on a half-infinite curve starting at
a point $z_0$ such that
the arc $\DD$ always starts at $z_0$ as the times independently
vary in some open set.

\section*{Acknowledgments}

The author thanks the organizers of the Workshop
``Physics and mathematics of growing interfaces"
(Santa Fe, January 2006) for a nice and stimulating meeting.
Discussions with Ar.Aba\-nov, E.Bet\-tel\-heim,
I.Kri\-che\-ver, M.Mi\-ne\-ev-\-Wein\-stein,
T.Ta\-ke\-be, L.P.-Teo and P.Wiegmann are gratefully acknowledged.
This work was supported in part
by grant INTAS 03-51-6346,
by grant for support of scientific schools
NSh-8004.2006.2 and by the ANR project GIMP No. ANR-05-BLAN-0029-01.

\end{document}